\documentclass[prb,cha,onecolumn,superscriptaddress,preprintnumbers,showpacs, amsmath,amssymb]{revtex4-1}
\usepackage{bm}
\usepackage{hyperref}
\hypersetup{colorlinks=true,linkcolor=blue,citecolor=blue}
\usepackage{amsmath}
\usepackage{amssymb}
\usepackage{amsthm}
\usepackage{amsfonts}
\usepackage{enumerate}
\usepackage{latexsym}
\usepackage{ifpdf}
\usepackage{graphicx}
\usepackage{makeidx}
\expandafter\ifx\csname package@font\endcsname\relax\else
\expandafter\expandafter
\expandafter\usepackage
\expandafter\expandafter
\expandafter{\csname package@font\endcsname}
\fi
\usepackage{color}
\usepackage{times}

\begin{document}

\title{The Rise of Refractory Transition-metal Nitride Films for Advanced Electronics and Plasmonics}

\author{Jiachang Bi}
\email{bijiachang@nimte.ac.cn}
\affiliation{Ningbo Institute of Materials Technology and Engineering, Chinese Academy of Sciences, Ningbo 315201, China}
\author{Ruyi Zhang}
\affiliation{Ningbo Institute of Materials Technology and Engineering, Chinese Academy of Sciences, Ningbo 315201, China}
\author{Xiong Yao}
\affiliation{Ningbo Institute of Materials Technology and Engineering, Chinese Academy of Sciences, Ningbo 315201, China}
\author{Yanwei Cao}
\email{ywcao@nimte.ac.cn}
\affiliation{Ningbo Institute of Materials Technology and Engineering, Chinese Academy of Sciences, Ningbo 315201, China}
\affiliation{Center of Materials Science and Optoelectronics Engineering, University of Chinese Academy of Sciences, Beijing 100049, China}

\date{\today}

\begin{abstract}

The advancement of semiconductor materials has played a crucial role in the
development of electronic and optical devices. However, scaling down
semiconductor devices to the nanoscale has imposed limitations on device
properties due to quantum effects. Hence, the search for successor materials
has become a central focus in the fields of materials science and physics.
Transition-metal nitrides (TMNs) are extraordinary materials known for their
outstanding stability, biocompatibility, and ability to integrate with
semiconductors. Over the past few decades, TMNs have been extensively
employed in various fields. However, the synthesis of single-crystal TMNs has
long been challenging, hindering the advancement of their high-performance
electronics and plasmonics. Fortunately, progress in film deposition
techniques has enabled the successful epitaxial growth of high-quality TMN
films. In comparison to reported reviews, there is a scarcity of reviews on
epitaxial TMN films from the perspective of materials physics and condensed
matter physics, particularly at the atomic level. Therefore, this review aims to
provide a brief summary of recent progress in epitaxial growth at atomic
precision, emergent physical properties (superconductivity, magnetism,
ferroelectricity, and plasmon), and advanced electronic and plasmonic devices
associated with epitaxial TMN films.

\end{abstract}


\maketitle
\newpage

\section{Introduction}

One generation of material powers one generation of devices. The successful synthesis of single crystalline semiconductors, such as silicon (Si), gallium nitride (GaN), and silicon carbide (SiC), has been fundamental to the development of semiconductor chips. Moore's law has dictated that the number of transistors in Si-integrated circuits doubles every two years. However, as transistor scaling continues to reach 1 nanometer, the limitations of their operating principles are becoming apparent. \cite{Nature-2015-More,Nature-2016-Moore,Nature-2019-Moore,Nature-2023-Moore} Therefore, surpassing Moore's law requires the exploration of alternative materials to conventional silicon, which has emerged as a focal point in the field of materials science and physics. 

Several potential electronic materials have been proposed, including transition metal oxides (TMOs) and transition metal dichalcogenides (TMDs). \cite{RMP-2014-Chakhalian,NM-2012-Chakhalian,ARMR-2016-Middey, NM-2012-Hwang,MRS-2008-Mannhart,NRM-2017-Martin,ARMR-2007-Schlom,JACS-2008-Schlom,MSEB-2001-Schlom,NM-2007-Ramesh,NRM-2019-Ramesh,PhyT-1998-Auciello,NM-2019-Spaldin,PhyT-2010-Spaldin,MaterS-2010-Martin,RMP-2005-Dawber,NC-2018-Kageyama,NRM-2017-Manzeli,NatNano-2012-Wang,RMP-2018-Wang,AM-2021-Fu,NChe-2013-Chhowalla,2DMater-2016-Lin}  Due to the strong interactions among charge, spin, orbital, and lattice degrees of freedom in transition-metal oxides, the electrons in these materials exhibit strong correlations, leading to the emergence of phenomena such as high-temperature superconductivity, colossal magnetoresistance, and metal-insulator transitions. \cite{RMP-2003-Cuprate,JPCM-1997-Colossal,RMP-1998-MIT}
Additionally, researchers have constructed intriguing devices using oxide films to exploit their exotic quantum states, \cite{NP-2011-LAO,NM-2008-MIT} signifying the advent of the oxide electronics era. \cite{Science-2007-Oxide} Recently, there has been significant interest in the study of two-dimensional materials, particularly transition-metal dichalcogenides, with a special focus on MoS$_2$.\cite{Nature-2015-MoS2,Nature-2018-MoS2,NE-2020-MoS2,Nature-2022-MoS} In fact, single monolayer MoS$_2$ transistors with sub-1-nm channel length have been successfully fabricated on a 2-inch wafer, highlighting the promising future of transition-metal dichalcogenides. \cite{Nature-2022-MoS} Furthermore, several novel  logic devices, including spin field-effect transistors requiring the formation of spin-polarized two-dimensional electron gases, have been proposed. \cite{APL-1990-Spin} However, the lack of electronic materials with effective spin-polarized electron gases presents a significant obstacle to the development of spin field-effect transistors. Beyond the operating principles of conventional logic devices, the introduction of quantum devices has revolutionized the foundations of logic devices. Notably, in 2019, Google announced the development of superconducting quantum chips comprising only 53 qubits, which exhibited unexpected power. \cite{Nature-2019-Quantum} Presently, the quest for an ideal material for quantum computing hardware, ranging from elementary metals like Al and Ta to transition metal nitrides such as titanium nitride (TiN), remains a considerable challenge. \cite{Science-2021-Material}

\begin{figure}[htpb]
	\centering
	\includegraphics[width=0.9\linewidth]{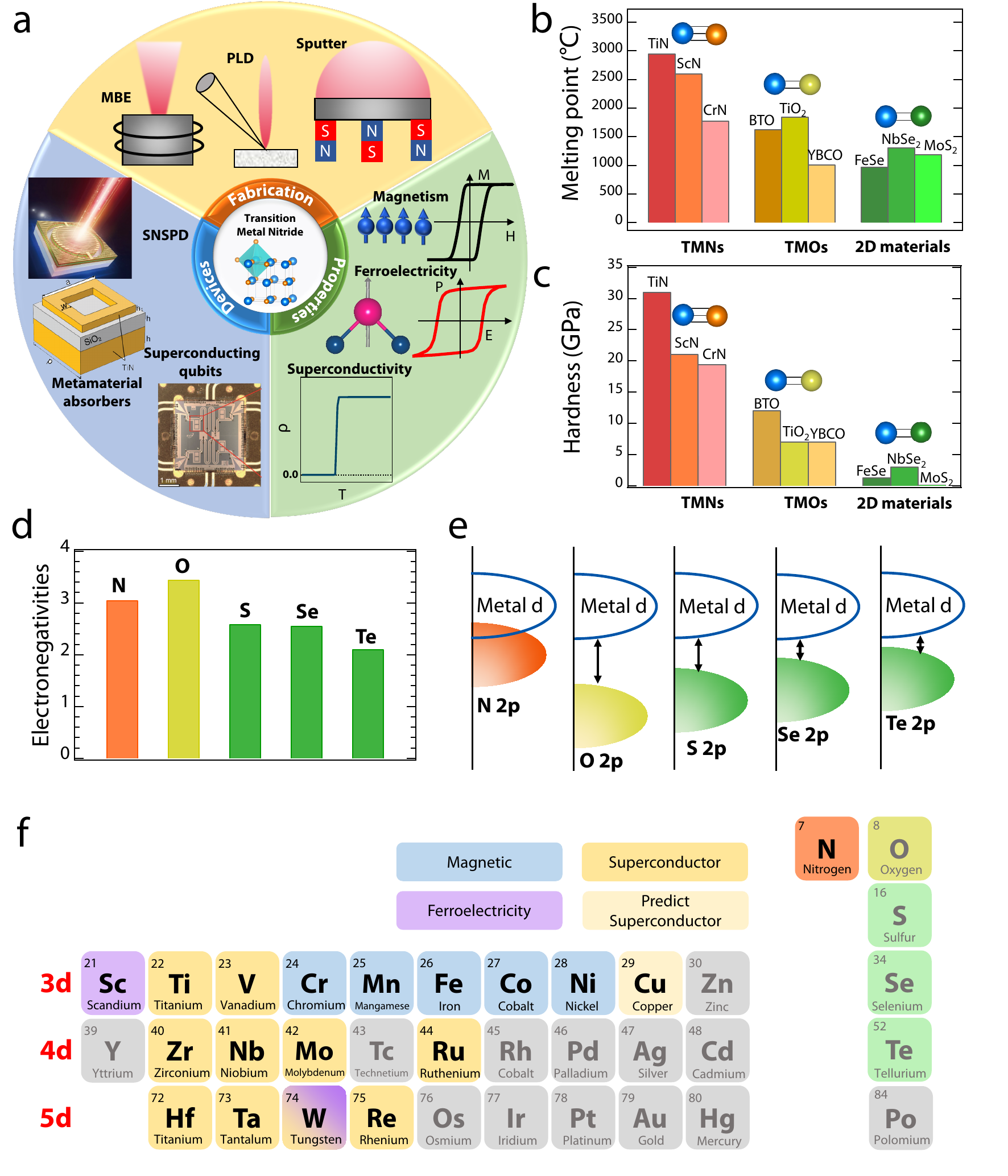}
	\caption{ a) The synthesis, properties, and devices of transition metal nitride films. \cite{PHYSICS-2021-SNSPD,AM-2014-Li,CM-2021-Qubit}  b) The melting point and c) hardness of transition metal nitrides (TiN, \cite{Book-1996-TiN,JVST-1991-TiNHard} ScN, \cite{JMS-2004-ScN,Materialia-2022-ScNHardness} and CrN \cite{JMPB-2017-CrNHardness,book-2012-CrN}), transition metal oxides (BaTiO$_3$, \cite{book-2016-BTO,JElectr-2008-BaTiO3Hardness} TiO$_2$, \cite{JAC-1952-TiO2Point,Book-2020-TiO2Hardness} and YBa$_2$Cu$_3$O$_{7-x}$ \cite{PhysicaC-2000-YBaCuOPoint,BullMater-2013-YBaCuOHardness}), and transition metal 2D materials (FeSe, \cite{JPED-1968-FeSe,PhysicaC-2010-FeSeHardness} NbSe$_2$, \cite{book-2011-NbSe,Friction-2023-NbSe2Hardness} and MoS$_2$ \cite{Nature-1959-MoS,Coatings-2019-MoS2Hardness}). d)  The electronegativities of  N, O, S, Se, and Te.\cite{NC-2018-Kageyama} e) The hybridization between transition-metal cations and N, O, S, Se, and Te anions. f) Superconductivity, magnetism, and ferroelectricity of TMNs.}
\end{figure}

Transition-metal nitrides (TMNs) are remarkable materials that exhibit physical properties with the characteristics of both metals and ceramics. As a result, they have found extensive applications in various fields, including electronics, plasmonics, catalysis, coatings, thermoelectrics, and biomedicine (see Figure 1a). \cite{AFM-2021-Photocatalytic,CSR-2021-Electrochemical, ChemCata-2024-Mou, GEE-2023-Luo,Catal-2023-Park,FM-2020-Ashraf,JES-2023-Kadam,PMS-2015-Ningthoujam, JMMC-2016-Eklund,AM-2013-BeyondGold,Science-2014-Refractory,AM-2014-Li,Science-2015-Gold,CM-2022-Dasog,ThinSolidFilms-1994-Steinmuller,TheSolidFilms-1985-Sundgren,JOM-2001-Navinsek,ActaMater-2004-Yang,JAP-1987-Helmersson,JJAP-2019-Review,APR-2018-Saha,PRM-2019-Biswas} For example, semiconducting ScN has a simple crystal structure but can exhibit impressive multifunctional performance, potentially enabling novel device functionalities. \cite{APR-2018-Saha,PRM-2019-Biswas} Robust superconductivity is another extraordinary property of TMNs. The investigation of superconductivity in TMNs, such as TiN, VN, and ZrN, can date back to the 1930s. \cite{ZP-1930-Nitride} In 1952, it was discovered that hexagonal MoN could display superconductivity at temperatures near 12 K, which represented the second-highest transition temperature at that time. \cite{PRB-1952-MoN} Subsequently, NbN, with a superconducting transition temperature near 17 K, was synthesized and widely utilized. \cite{JAP-1971-NbNTc,JAP-1967-NbTiN,CM-2021-Qubit,PHYSICS-2021-SNSPD,PRA-2019-SCWR} In recent years, magnetic and ferroelectric properties have also been observed in TMNs. \cite{JAP-2019-Ferr-AlScN,Science-2021-LaWN}  Apart from TMOs and TMDs, TMNs possess exceptional stability in terms of their chemical, mechanical, and thermal properties (see Figure 1b-d). \cite{Book-1996-TiN,JVST-1991-TiNHard,JMS-2004-ScN,Materialia-2022-ScNHardness,JMPB-2017-CrNHardness,book-2012-CrN,book-2016-BTO,JElectr-2008-BaTiO3Hardness,JAC-1952-TiO2Point,Book-2020-TiO2Hardness,PhysicaC-2000-YBaCuOPoint,BullMater-2013-YBaCuOHardness,JPED-1968-FeSe,PhysicaC-2010-FeSeHardness,book-2011-NbSe,Friction-2023-NbSe2Hardness,Nature-1959-MoS,Coatings-2019-MoS2Hardness} Interestingly, the melting point of TMNs can reach up to 3000$^\circ$C, and they can exhibit a hardness of 30 GPa. \cite{Book-1996-TiN,JVST-1991-TiNHard} As a result, TMNs are considered refractory and hard superconductors. \cite{PNAS-2005-HSC} It is worth noting that most TMNs are metallic, whereas most TMOs and TMDs are insulating and semiconducting, resulting from the strong hybridization between transition metal and nitrogen ions (see Figure 1e). These exceptional properties make multifunctional TMNs (see Figure 1f) advantageous for the development of advanced electronic and plasmonic devices.

However, the synthesis of high-quality epitaxial transition-metal nitrides at the atomic level has long been a challenge due to their refractory nature, their complex phase diagram, and the inactive characteristic of nitrogen molecules. Therefore, only small TiN crystals can be synthesized at very high temperatures (above 2000 $^\circ$C),\cite{PRB-1980-Johansson,JCG-1976-Motojima} and there are numerous defects in the early-prepared single crystalline TiN films,\cite{JAP-1987-Helmersson,ThinSolidFilm-1992-Hultman} which has hindered advancements in high-performance electronics and plasmonics. Compared to a large number of great reviews on high-quality transition metal oxide films and transition metal dichalcogenides,\cite{RMP-2014-Chakhalian,NM-2012-Chakhalian,ARMR-2016-Middey, NM-2012-Hwang,MRS-2008-Mannhart,NRM-2017-Martin,ARMR-2007-Schlom,JACS-2008-Schlom,MSEB-2001-Schlom,NM-2007-Ramesh,NRM-2019-Ramesh,PhyT-1998-Auciello,NM-2019-Spaldin,PhyT-2010-Spaldin,MaterS-2010-Martin,RMP-2005-Dawber,NC-2018-Kageyama,NRM-2017-Manzeli,NatNano-2012-Wang,RMP-2018-Wang,AM-2021-Fu,NChe-2013-Chhowalla,2DMater-2016-Lin} there is a scarcity of reviews on high-quality epitaxial TMN films at the atomic scale, especially from the perspective of materials physics and condensed matter physics.  

In this review, we offer a brief overview of recent advancements in epitaxial transition-metal nitride films from the perspective of materials physics and condensed matter physics.  Advanced deposition techniques, such as MBE, PLD, and magnetron sputtering epitaxy, capable of synthesizing high-quality epitaxial transition-metal nitride films at the atomic scale were introduced. We focus on summarizing their physical properties, e.g., superconductivity, magnetism, ferroelectricity, and plasmon, as well as their applications in advanced electronic and plasmonic devices.

\section{Epitaxial Growth}

The epitaxial growth of TMN films poses significant challenges for several reasons. Firstly, unlike active oxygen gas, nitrogen molecules are inactive and require nitrogen plasma assistance for the reaction of transition metals. Moreover, it is important to note that the bond energy of N $\equiv $ N (around 945 kJ/mol) is considerably higher than that of O = O (around 498 kJ/mol).\cite{Handbook-2005-Lide} Therefore, the breaking of nitrogen molecules requires more energy. Secondly, most TMNs have high melting points, reaching up to 3000$^\circ$C. Hence, to synthesize high-quality films, the growth temperature for TMN films is typically high. While there are some reports on low-temperature (nearly room temperature) epitaxy of TMN films, \cite{PRM-2022-Gao,JVST-2023-Rao,JVST-2018-Villamayor,OptMater-2015-Yang,OptMater-2015-Zgrabik,SurCoatTech-2007-Yang,JJAP-2003-Chen,JAP-2003-Chen,ThinSolidFilms-2010-Lee,CryDes-2017-Rasic,ACSPhoto-2018-Sugavaneshwar,APL-2016-Briggs} their crystalline quality is generally worse than that of films grown at high temperatures in most cases. \cite{ACSPho-2019-Guo,JAP-2012-Krockenberger,JAP-2020-Richardson, PRM-2021-Bi,ACSPho-2021-Zhang,ACSInter-2021-Zhang,NanoLett-2023-Zhang,AIP-2020-Peng} Thirdly, the phase diagrams of TMNs are more complex compared to common nitrides like GaN, due to the presence of multiple valence states of transition metal ions. \cite{Book-2015-Lengauer,Book-2014-Khomskii} Consequently, in contrast to extensively studied TMOs and TMDs, the epitaxial growth of TMN films remains relatively uncommon. In addition to well-known methods like molecular beam epitaxy (MBE) and pulsed laser deposition (PLD), \cite{JACS-2008-Schlom,MSEB-2001-Schlom,NM-2007-Ramesh, JPCM-2008-Christen,Science-1996-Lowndes,Book-2007-Eason,Book-2017-Biswas} we emphasize a high-pressure magnetron sputtering epitaxy growth technique (see Figure 2c), \cite{PRM-2021-Bi,ACSPho-2021-Zhang,ACSInter-2021-Zhang,NanoLett-2023-Zhang,AIP-2020-Peng,APLMater-2021-Zhang,AIPAdvances-2021-Bi,CPL-2023-Li,APL-2024-Hou} to synthesize the high-quality epitaxial transition metal nitride films at the atomic level. 

\begin{figure}[htpb]
	\centering
	\includegraphics[width=0.95\linewidth]{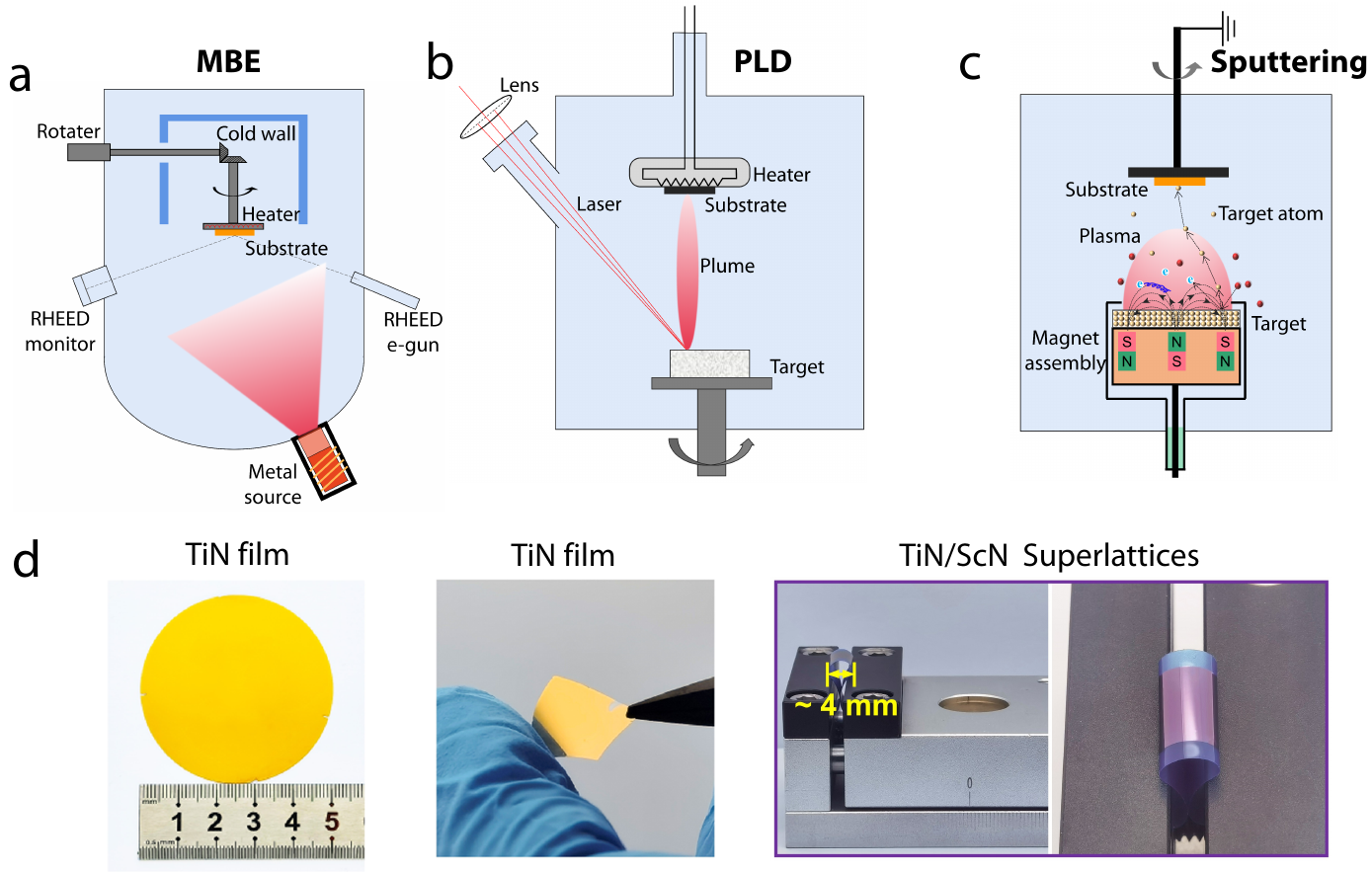}
	\caption{ Schematic of a) MBE, b) PLD, and c) Sputtering. d) Photos of TiN films \cite{ACSInter-2021-Zhang} and TiN/ScN/TiN/ScN  superlattices. \cite{NanoLett-2023-Zhang} Reproduced with permission.\cite{ACSInter-2021-Zhang} Copyright 2021, American Chemical Society. Reproduced with permission.\cite{NanoLett-2023-Zhang}  Copyright 2023, American Chemical Society.}
\end{figure} 

MBE and chemical vapour deposition (CVD) are widely recognized as the standard for ultra high-quality epitaxial thin film growth, particularly for epitaxial semiconductor films. A prime example of its capabilities is the attainment of an ultra-high mobility two-dimensional electron gas (5.7 $\times$10$^7$ cm$^2$/V·s) in high-quality AlGaAs/GaAs heterostructures grown by MBE. \cite{PRB-2022-Chung} MBE entails the generation of low-energy ($\sim$1 eV) thermal atomic beams, which can be achieved using Knudsen effusion cells or electron beam evaporators (see Figure 2a). To overcome the high kinetic barriers of N$_2$ dissociation and activate the N$_2$ source at a moderate temperature, radio frequency (RF) plasma-assisted MBE (PAMBE or RFMBE) is commonly utilized to break the nitrogen bonds during nitride film growth,\cite{CryGrow-2001-NitrideMBE} and the nitrogen plasma technology is mature. By employing a plasma source in the plasma chamber, molecular nitrogen gas can be dissociated into atomic nitrogen while maintaining a pressure range of 10$^{-3}$ Torr, whereas the pressure in the growth chamber can be maintained at 10$^{-5}$ to 10$^{-7}$ Torr during film growth. In some cases, ammonia (NH$_3$-MBE) has also served as a nitrogen source for nitride film growth. \cite{SurSci-2001-Grachev} RF plasma-assisted MBE has demonstrated success in the growth of various epitaxial TMN thin films, including superconducting TiN, NbN, ferroelectric Sc$_{0.18}$Al$_{0.82}$N, magnetic Fe-N, Mn$_3$N$_2$, Mn$_4$N, and CrN films (see Figure 3). \cite{ACSPho-2019-Guo,JAP-2012-Krockenberger,JAP-2020-Richardson,PRM-2023-Wright,PRM-2021-Wright,APLMater-2022-Wright,APL-2001-Yang,APL-2018-Meng,MaterResB-2018-Li,APL-2006-Ney,AEM-2022-AlScN,Nature-2018-AlN-NbN, NanoLett-2019-Mn4N}

 \begin{figure}[t]
 	\centering
 	\includegraphics[width=1\linewidth]{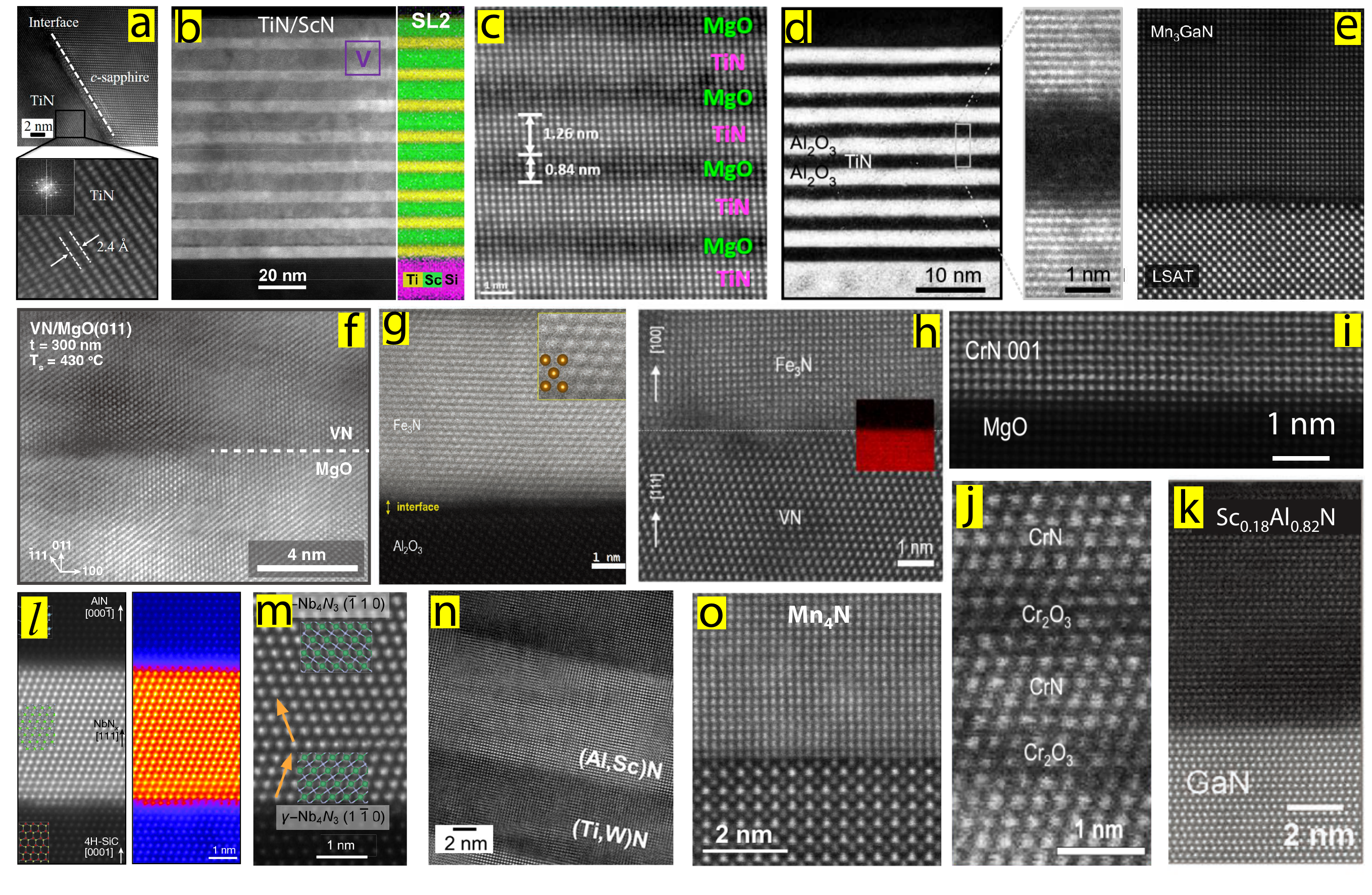}
 	\caption{Transmission electron microscopy images of  a) TiN films grown by MBE, \cite{ACSPho-2019-Guo} Reproduced with permission.\cite{ACSPho-2019-Guo} Copyright 2019, American Chemical Society. b) TiN/ScN superlattices grown by sputtering, \cite{NanoLett-2023-Zhang} Reproduced with permission.\cite{NanoLett-2023-Zhang} Copyright 2023, American Chemical Society. c) TiN/MgO superlattices grown by PLD, \cite{MaterToday-2021-Huang} Reproduced with permission.\cite{MaterToday-2021-Huang} Copyright 2021, Elsevier. d) TiN/Al$_2$O$_3$ superlattices grown by sputtering, \cite{SciAdv-2020-Quantum} Reproduced with permission.\cite{SciAdv-2020-Quantum} Copyright 2020, American Association for the Advancement of Science. e) MnGa$_3$N films grown by sputtering, \cite{NC-2020-MnGaN} Reproduced with permission.\cite{NC-2020-MnGaN} Copyright 2020, Nature Publishing Group UK London. f) VN films grown by sputtering, \cite{PRB-2015-VN}  Reproduced with permission.\cite{PRB-2015-VN} Copyright 2015, American Physical Society. g) Fe$_3$N films grown by PLD, \cite{AM-2023-Fe3N} Reproduced with permission.\cite{AM-2023-Fe3N}  Copyright 2023, Wiley-VCH. h) Fe$_3$N/VN heterostructures grown by PLD, \cite{NSR-2023-FeN-VN} Reproduced with permission.\cite{NSR-2023-FeN-VN}  Copyright 2024, Oxford University Press. i) CrN films grown by PLD, \cite{PRM-2021-Jin} Reproduced with permission.\cite{PRM-2021-Jin} Copyright 2021, American Physical Society. j) CrN/Cr$_2$O$_3$/CrN/Cr$_2$O$_3$ superlattices grown by PLD, \cite{PRL-2022-Jin} Reproduced with permission.\cite{PRL-2022-Jin} Copyright 2022, American Physical Society. k) Sc$_{0.18}$Al$_{0.82}$N films grown by MBE, \cite{AEM-2022-AlScN} Reproduced with permission.\cite{AEM-2022-AlScN} Copyright 2022, Wiley-VCH. l) AlN/NbN heterostructures grown by MBE, \cite{Nature-2018-AlN-NbN} Reproduced with permission.\cite{Nature-2018-AlN-NbN} Copyright 2018, Nature Publishing Group UK London. m) $\gamma$-Nb$_4$N$_3$ films grown by MBE, \cite{PRM-2021-Wright} Reproduced with permission.\cite{PRM-2021-Wright} Copyright 2021, American Physical Society. n) (Ti,W)N/(Al,Sc)N superlattices grown by sputtering,\cite{PRB-2016-Saha} Reproduced with permission.\cite{PRB-2016-Saha}  Copyright 2016,  American Physical Society. and o) Mn$_4$N films grown by MBE, \cite{NanoLett-2019-Mn4N} Reproduced with permission.\cite{NanoLett-2019-Mn4N} Copyright 2019, American Chemical Society. } 
 \end{figure}
In academic research, PLD is one of the most commonly used technique for growing high-quality epitaxial oxide films, particularly complex oxide films such as high-temperature superconducting YBa$_2$Cu$_3$O$_{7-x}$ films. \cite{JPCM-2008-Christen,Science-1996-Lowndes,Book-2007-Eason,Book-2017-Biswas,APL-1987-Dijkkamp} The schematic of PLD is shown in Figure 2b. This method involves utilizing a laser, typically a KrF laser with a wavelength of 248 nm or a frequency-doubled Nd:YAG laser with a wavelength of 532 nm, as an external energy source to ablate an oxide target,  resulting in the generation of a highly energetic plasma plume. \cite{Book-2017-Biswas} This plasma plume consists of ions and molecules that are subsequently deposited onto the substrate surface. For oxide films, oxygen is chemically active and does not require external activation treatment. Conversely, for the growth of nitride films, the nitrogen molecule is chemically stable, and the nitrogen-nitrogen triple bond is too stable to ionize effectively. Therefore, like MBE, an external RF nitrogen atom source is also required for the growth of TMN films via MBE and PLD. PLD has been successfully used to grow superconducting TiN, VN, ZrN, and magnetic CrN, Fe$_3$N films  (see Figure 3). \cite{ACSPhoto-2018-Sugavaneshwar,MaterToday-2021-Huang,PRB-2022-Zou,SciBul-2023-Chen,PRL-2022-Jin,AM-2021-Jin,PRM-2021-Jin,IOP-2018-Torgovkin,AM-2023-Fe3N,NSR-2023-FeN-VN} However, due to the limited wavelengths of the laser, which is hard to interact with metals, PLD targets are typically powder-sintered ceramics. Usually, the purity of ceramic targets is less than 99.95\%, notably lower than the purity of pure metal targets (99.99\%-99.9999\%).

Sputtering is a highly suitable technology for synthesizing TMN films, especially for depositing high-melting-point materials. The schematic of the sputtering system is shown in Figure 2c. In previous reports on magnetron sputtering epitaxy, there is an unusual setup (90 degrees off-axis sputtering) between the sputtering gun and the sample holder.\cite{Sccience-1990-Eom,ACSAMI-2021-Posadas,APL-2013-Peters} This method can effectively deliver the source material from the target. \cite{IOP-2020-Kobayashi} Interestingly, during the sputtering process, the argon or nitrogen plasma is naturally generated by applying a high voltage between the cathode (the assembly of the sputtering target and the copper holder,safety note: never have it connected to electricity without a target present and mounted on the chamber, with protective metal cap!) and the anode (grounded chamber and grounded other parts). \cite{IOP-2020-Gudmundsson} Therefore, an external RF nitrogen atom source is not required for sputtering. Within the plasma, high-energy ions bombard the target surface, leading to the deposition of films on substrates. \cite{IOP-2020-Gudmundsson,MaterS-2010-Martin} Hence, there are several advantages of magnetron sputtering for TMN film growth. First, sputtering can effectively ionize nitrogen by the sputtering gun without adding an external RF nitrogen atom source. However, both PLD and MBE rely on an external RF nitrogen atom source, where the ionization efficiency of nitrogen is low, resulting in nitrogen-deficient TMNs films. Secondly, sputtering can grow large-size epitaxial films (see Figure 2d), which is crucial for device applications. The growth of large-size single-crystal epitaxial films is challenging with PLD due to the small laser spot, whereas MBE equipment, capable of growing large-size single-crystal epitaxial thin films, is very expensive. Thirdly, sputtering can grow highly pure thin films with high-purity metal targets and nitrogen, which is crucial for superconductors and devices.\cite{MaterS-2010-Martin} In contrast, the limited wavelengths of the laser in PLD make it difficult to interact with metals, resulting in lower-purity TMNs films grown by PLD. Finally, magnetron sputtering epitaxy stands out as a versatile deposition technique capable of synthesizing various types of films (metals, insulators, semiconductors, and more). In contrast, MBE falls short in growing thin films like tungsten nitride due to tungsten's high melting point. Although sputtering has been employed for synthesizing various TMN films, \cite{PRM-2021-Bi,ACSPho-2021-Zhang,ACSInter-2021-Zhang,NanoLett-2023-Zhang,PRB-2004-TiN,IOP-2020-Kobayashi,PRB-2004-TiN,JAP-2004-TiNx,IEEE-2015-TiNSi,AOM-2017-UltrathinTiN,PRM-2017-VN,JMCC-2016-VNx,PRB-2015-VN,IOP-2020-Kobayashi,PPB-2006-FeN,PRB-2011-CrN,APL-2015-MnN,SciAdv-2020-Quantum,NC-2020-MnGaN,PRB-2016-Saha,PRL-2023-Biswas} achieving high-quality epitaxial growth of TMN films at the atomic level through sputtering is still a challenging work (see Figure 3).

\section{Physical Properties}

\subsection{Superconductivity}

Although the investigation of superconductivity in TMNs dates back to the 1930s, \cite{ZP-1930-Nitride} the nature of superconductivity in TMNs remains a topic of ongoing debate. As shown in Figure 4a, reported superconducting transition temperatures (T$\rm_c$) vary across different TMNs: 5.6 K / 6K for TiN, \cite{PRB-1978-TiN6K,JLCM-1978-TiNPressure} 9.2 K for VN, \cite{PRB-1984-PressureVN} 10 K for ZrN, 17 K for NbN, \cite{JAP-1971-NbNTc} 14 K for MoN,\cite{JMST-1987-MoN,CM-2008-MoN} 8.8 K for HfN, \cite{Book-1996-TiN} 10.8 K for TaN, 4.85 K for W$_2$N, \cite{JAP-1978-TaN,JVST-1975-WN} and 5 K for ReN$_x$ (see Figure 4a). \cite{JLTP-1983-ReN,JAC-2009-ReN} Recently, superconducting properties have been discovered in RuN with T$\rm_c$ ranging from 0.77 K to 1.29 K. \cite{AirXiv-2023-RuN} Theoretical predictions also propose that CuN could potentially exhibit a high T$\rm_c$ of up to 30 K, attributed to a strong electron-phonon coupling strength. \cite{PRB-2023-CuN}

For conventional superconductors, the superconducting properties can be well described by the BCS theory. However, the density of states at the Fermi surface \cite{PRL-2001-HfN} and the electron–phonon coupling strength \cite {PhysicaC-2015-Layered} of TMNs are too low to account for their relatively high T$\rm_c$,  extending beyond the BCS framework. In the case of TiN, the highest reported T$\rm_c$  is 6 K for bulk TiN \cite{PRB-1978-TiN6K,JLCM-1978-TiNPressure} and 5.6 K for TiN films.\cite{PRA-2019-Disorder} Interestingly, disordered TiN films can exhibit various exotic physical phenomena. For example, with increasing disorder, the TiN films can transition from superconducting to insulating states.\cite{PRL-2007-Disorder,PRL-2008-Disorder,JETP-2004-SIT,PhysicaB-2005-SIT} The electronic texture in these films likely comprises an array of small superconducting islands coupled by Josephson weak links. \cite{Book-2004-SC}  The presence of disorder enforces a droplet-like electronic texture, where superconducting islands immerse into a normal matrix, and tuning the disorder can drive the system from a superconducting to an insulating state, forming the Cooper-pair insulator. \cite{Nature-2008-Superinsulator}  Near the superconductor-insulator transition, a Cooper-pair insulator emerges. Notably, this insulator shows an infinite resistance, exhibiting a duality to the superconducting state, as shown in Figure 4b. Interestingly, pseudogap exists in ultrathin TiN films over a wide range of temperatures above T$\rm_c$ (see Figure 4c), \cite{NC-2010-Pseudogap} which is a characteristic often associated with high-temperature superconductivity. The pseudogap state in disordered TiN films can be induced by superconducting fluctuations and be favoured by two-dimensionality. \cite{NC-2010-Pseudogap} A remarkable experimental finding was the formation of Cooper pairs above T$\rm_c$ even in the absence of a spectroscopic (pseudo) gap. \cite{Science-2021-CooperPairing} The spectroscopic gap in this case fills up rather than closes with increasing temperature, indicating the existence of a state above T$\rm_c$ that has no (pseudo)gap but carries charge through paired electrons,\cite{Science-2021-CooperPairing} as shown in Figure 4d. 

 \begin{figure}[htpb]
	\centering
	\includegraphics[width=0.88\linewidth]{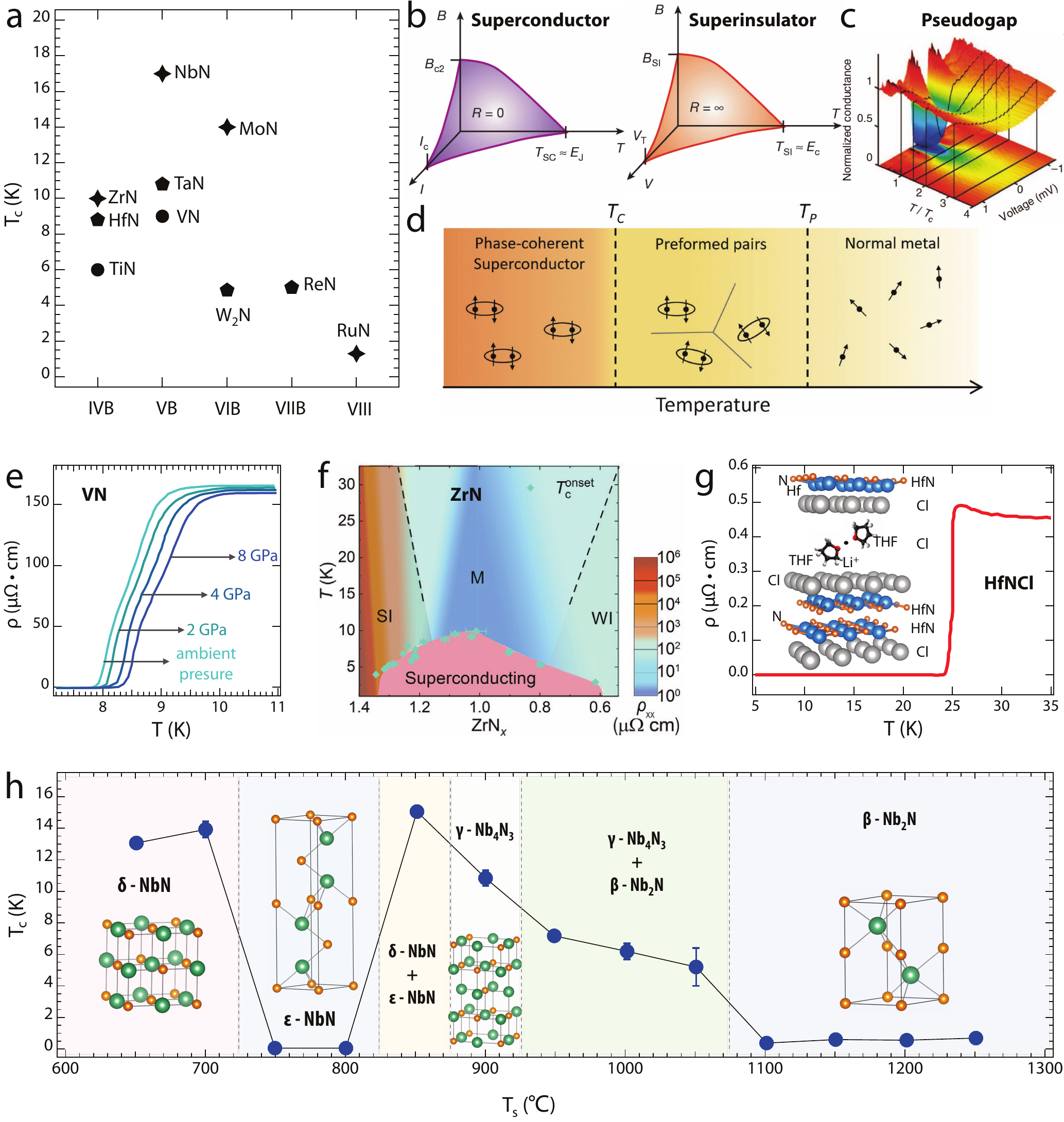}
	\caption{ a) The summary of T$\rm_c$ values in TMNs. b) Sketch of dual-phase diagrams for a superconductor and a superinsulator.  \cite{Nature-2008-Superinsulator} Reproduced with permission.\cite{Nature-2008-Superinsulator} Copyright 2008, Nature Publishing Group UK London. c) and d) Tunnelling conductance and temperature-dependent states of Cooper pairs in TiN films. \cite{NC-2010-Pseudogap,Science-2021-CooperPairing} Reproduced with permission.\cite{NC-2010-Pseudogap} Copyright 2010, Nature Publishing Group UK London. Reproduced with permission.\cite{Science-2021-CooperPairing} Copyright 2021, American Association for the Advancement of Science. e) Pressure-dependent T$\rm_c$ of VN. \cite{JPSJ-2015-PressureVN} Reproduced with permission.\cite{JPSJ-2015-PressureVN} Copyright 2015, The Physical Society of Japan. f) Phase diagram of ZrN$_x$ films. \cite{SciBul-2023-Chen} Reproduced with permission. \cite{SciBul-2023-Chen}  Copyright 2023, Elsevier.  g) Temperature-dependent resistivity of Li$_{0.48}$(THF)$_y$HfNCl. The inset shows a schematic structural model of HfNCl layers.  \cite{Nature-1998-HfNCl} Reproduced with permission.\cite{Nature-1998-HfNCl}   Copyright 1998, Nature Publishing Group UK London. h) Phase diagram of NbN$_x$ films. \cite{PRM-2023-Wright} Reproduced with permission. \cite{PRM-2023-Wright} Copyright 2023, American Physical Society.}
\end{figure}

Also, in contrast to the conventional BCS-type superconductors, the T$\rm_c$ of VN actually increases with pressure (see Figure 4e),\cite{JPSJ-2015-PressureVN} reaching a maximum value at a certain pressure before decreasing, coinciding with the complete suppression of spin fluctuations. \cite{PRB-1980-Spin,JPCM-1992-VN} The highest reported T$\rm_c$ of VN films is near 9 K,\cite{PRB-1984-PressureVN,PRB-1985-VNfilms} while theoretically, T$\rm_c$ could reach 32 K in the absence of robust spin fluctuations.\cite{JPCS-1975-Magnetic,PRB-1980-Spin} Therefore, higher pressure is required to further explore the superconductivity of VN. \cite{JPCS-1975-Magnetic,JPSJ-1976-SF}  Moreover, the phase diagram of ZrN$_x$ exhibits  a notable similarity to that of high-T$\rm_c$ copper oxide superconductors. \cite{SciBul-2023-Chen} As shown in Figure 4f, a superconducting dome emerges in close proximity to a strongly insulating (SI) phase, with a metallic (M) state persisting above T$\rm_c$ near the optimal nitrogen concentration. Importantly, the electron-doped layered hafnium nitride $\beta$-HfNCl demonstrates a high T$\rm_c$ of up to 25.5 K (see Figure 4g, compared to the T$\rm_c$ of rock-salt HfN at $\sim$ 8.8 K\cite{Book-1996-TiN}).\cite{Nature-1998-HfNCl}  This feature is similar to the high-temperature superconductors based on copper oxides, where superconductivity occurs within two-dimensional CuO$_2$ planes separated by charge-reservoir oxide layers. \cite{Science-1990-Copper}  Additionally, under high pressures, dense Th$_3$P$_4$-type phases of Hf$_3$N$_4$ ($I\bar{4}3d$) can be formed, featuring higher coordination numbers of atoms, elevated elastic moduli, and possibly increased hardness. \cite{NM-2003-Hf3N4,PRL-2003-Hf3N4}

\begin{figure}[b]
	\centering
	\includegraphics[width=0.98\linewidth]{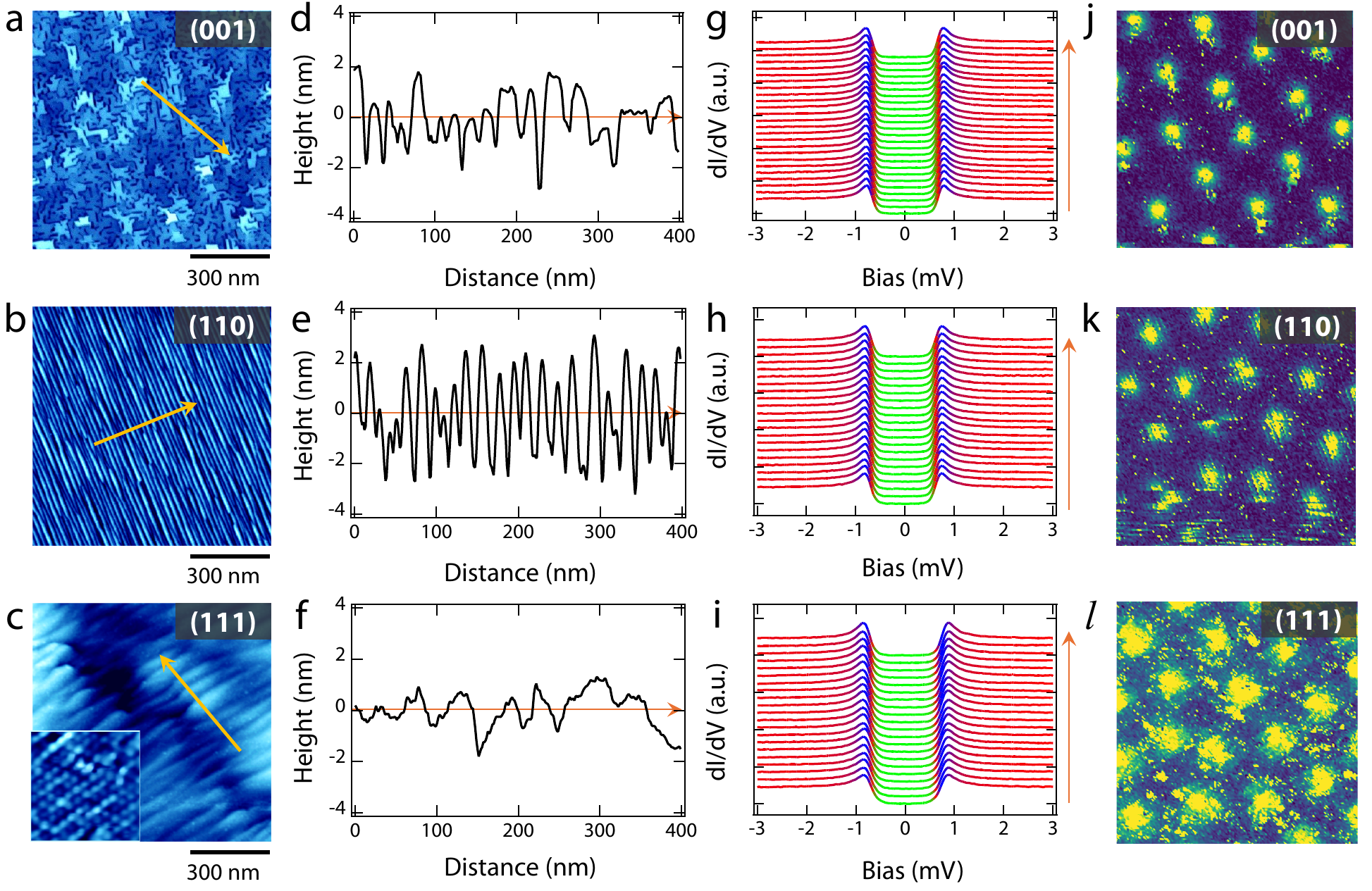}
	\caption{ a)-c) STM images of TiN (001), (110), and (111) films, tunneling conditions are V$\rm_{set}$ = $-$100 mV, I$\rm_{set}$ = 20 pA;  V$\rm_{set}$ = $-$100 mV, I$\rm_{set}$ = 20 pA; and V$\rm_{set}$ = $-$300 mV, I$\rm_{set}$ = 10 pA, respectively. The inset in (c) shows the atomic resolution on the TiN (111) surface with the size of 3.2 nm $\times$ 3.2 nm (V$\rm_{set}$ = $-$90 mV, I$\rm_{set}$ = 20 pA). d)-f) Height profile along the arrows marked in a)-c) showing the apparent roughness of the three films. g)-i) dI/dV spectra taken along the arrows marked in a)–c). j)-$l$) Images of the Abrikosov vortex lattices on different surfaces under a vertical magnetic field of 0.05 T.\cite{APL-2024-Zhang} Reproduced with permission.\cite{APL-2024-Zhang} Copyright 2024, AIP Publishing. }
\end{figure} 
In fact, the phase structure of TMNs is very complex,\cite{Book-2015-Lengauer,PRB-2017-Zhang} taking Nb-N as an illustrative example. Extensive studies have been conducted on its phase structure,\cite{PRM-2021-Wright,PRM-2023-Wright,JAP-1974-Oya,PRB-1995-Treece,PRB-2019-Babu,PRB-2023-Babu,Acta-2000-Lengauer,AMI-2022-Kobayashi} and the T$\rm_c$ of Nb-N films is highly dependent on it (see Figure 4h),\cite{PRM-2021-Wright,PRM-2023-Wright} which can be controlled by the growth parameters such as substrate temperature and active nitrogen flux.\cite{PRM-2023-Wright} NbN exhibits a notably high critical temperature (T$\rm_c$) of $\sim$17 K, surpassing other binary TMNs,\cite{JAP-1971-NbNTc,SST-2016-Hazra} and has been utilized in a wide variety of superconducting electronic devices.\cite{CM-2021-Qubit,PHYSICS-2021-SNSPD,PRA-2019-SCWR} The crystal structure of NbN yielding the highest T$\rm_c$ is currently a subject of debate.\cite{PRM-2023-Wright} The rock-salt structure $\delta$-NbN was thought to have the highest T$\rm_c$ but was unstable at room temperature.\cite{Acta-2000-Lengauer,SIA-1990-Lengauer,JLCM-1960-Brauer,PRB-2010-Ivashchenko} However, other reports indicated that the highest T$\rm_c$ exists in the mixed phases $\delta$-NbN and $\varepsilon$-NbN.\cite{PRM-2023-Wright,JVST-1980-Wolf,PCS-1988-Haase} Furthermore, the close-packed-hexagonal $\beta$-Nb$_2$N presents an opportunity for integration with the III-N semiconductor films. Therefore, further investigation is warranted to unravel the enigmatic superconducting properties of TMNs.

Recently, the microscopic morphologies and local density of states of high-quality superconducting TiN films were investigated by scanning tunneling microscope (STM). \cite{APL-2024-Zhang} Figure 5a–c depict typical topographic images for the TiN (001), (110), and (111) surfaces. The TiN (001) surface exhibits irregular fragments, whereas the TiN (110) surface looks like close-packed pine needles, and terraces are observed on the TiN (111) surface. There are numerous small grains with a height of $\sim$ 4 nm, as shown in Figure 5d,e. Notably, the TiN (111) surface consists of small flat areas with atomic resolution, as displayed in the inset in Figure 5c and 5f . The well-defined morphological features, especially the acquisition of atomic resolution, not only indicate the samples are of high quality and stability/high inertness, but also imply the electronic properties may be different in the three different orientated films. Further, scanning tunneling spectroscopy (STS) measurements were conducted to detect the superconducting states on these surfaces with grains or terraces. Figure 5g–i plot three sets of tunneling spectra measured along the lines marked in Figure 5a–c, respectively.\cite{APL-2024-Zhang} Remarkably, the spectra are spatially homogeneous and robust against the grains and terraces, which resembles the case in some dirty superconductors.\cite{CPB-2021-Zhu} In Figure 5j–$l$, zero energy STS maps exhibit ordered vortex lattices in all three films, confirming that TiN are type-II superconductors. It is noted that the superconductivity is uniform between the surface and bulk of TiN films, which is confirmed by the STM/STS results and transport measurements, respectively.\cite{APL-2024-Zhang} Moreover, the TiN films were ex-situ probed by the STM/STS. The robust superconductivity and well-defined surface morphology confirm the high quality and chemical stability of TiN films, which is essential to fabricate devices based on superconducting films.

\subsection{Magnetism}

Compared to other magnetic materials, magnetic TMNs possess several peculiar advantages, such as thermal stability, corrosion resistance, and cost-effectiveness. In this part, we mostly discuss ferromagnetic (FM), antiferromagnetic (AFM), and ferrimagnetic phases in binary TMNs compounds, while also covering some recently emerged ternary compounds. Figure 6a illustrates the various phases of magnetic TMNs depending on the nitrogen content.

\begin{figure}[t]
	\centering
	\includegraphics[width=0.9\linewidth]{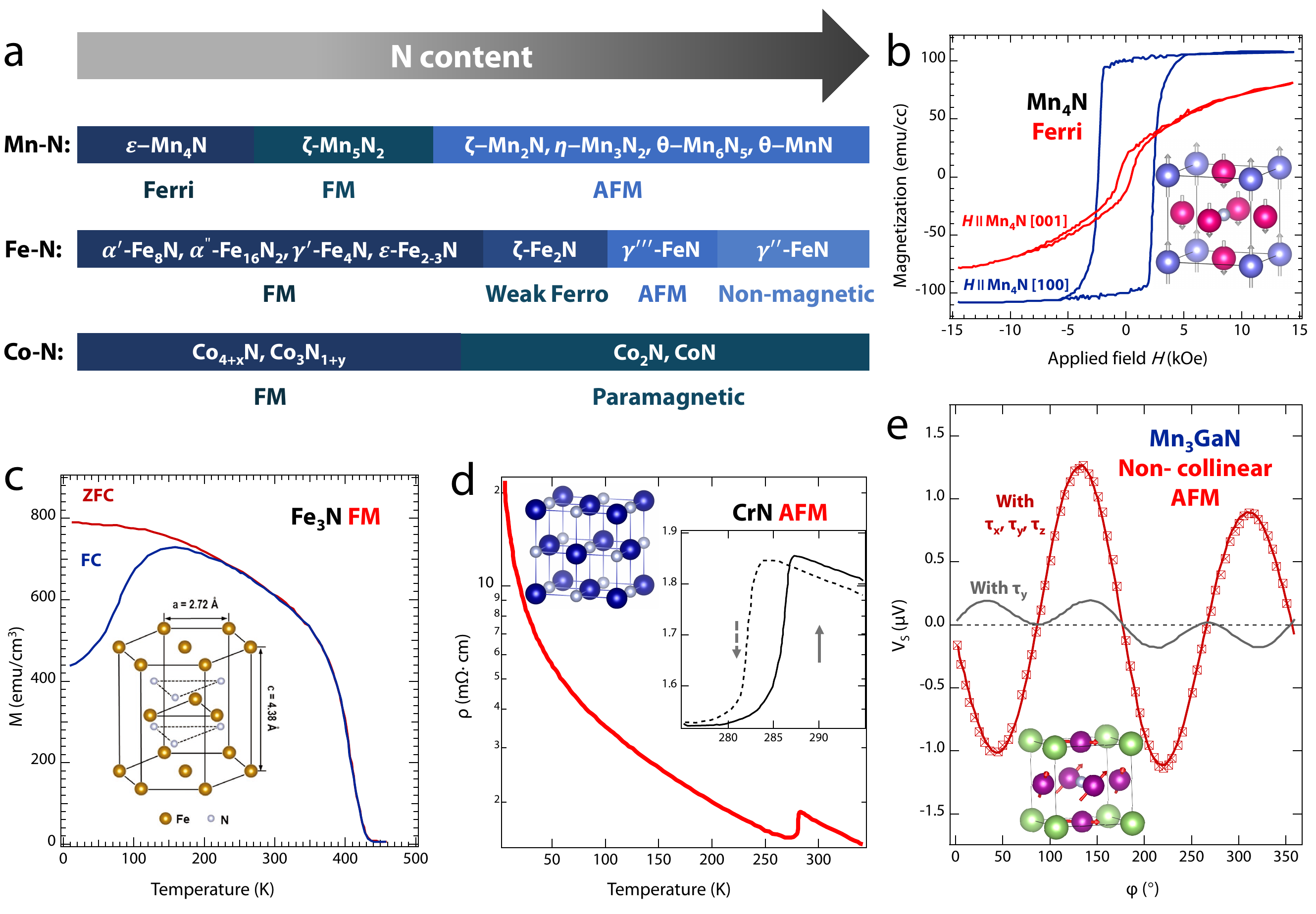}
	\caption{ a) Overview of nitrogen-content-dependent magnetic properties of various binary TMNs. b) Schematic of antiperovskite structure and magnetization curves of Mn$_4$N films. \cite{IOP-2022-Mn4N} Reproduced with permission.\cite{IOP-2022-Mn4N} Copyright 2022, IOP Publishing. c)Temperature-dependent magnetization and crystal structure of Fe$_3$N films. \cite{AM-2023-Fe3N} Reproduced with permission.\cite{AM-2023-Fe3N} Copyright 2023, Wiley-VCH. d) Temperature-dependent resistivity and crystal structure of CrN. \cite{PRL-2010-CrN} Reproduced with permission.\cite{PRL-2010-CrN}  Copyright 2010, American Physical Society. e) Symmetric spintorque ferromagnetic resonance components for the Mn$_3$GaN device as a function of the in-plane magnetic field angle and  antiperovskite structure of Mn$_3$GaN. \cite{NC-2020-MnGaN} Reproduced with permission.\cite{NC-2020-MnGaN} Copyright 2020, Nature Publishing Group UK London.}
\end{figure} 

The Mn-N family includes various compounds such as MnN, Mn$_6$N$_5$, Mn$_2$N, Mn$_3$N$_2$, Mn$_5$N$_2$, and Mn$_4$N. \cite{IOP-2022-Mn4N,JAC-2000-MnN} These compounds can be divided into four stable intermediate phases in the phase diagram, i.e., $\varepsilon, \zeta, \eta$, and $\theta$ phases. \cite{IOP-2022-Mn4N} MnN and Mn$_6$N$_5$ belong to the $\theta$ phase with the tetragonal structure, \cite{JAC-2000-MnN} whereas Mn$_2$N and Mn$_5$N$_2$ form the $\zeta$ phase with a hexagonal structure. \cite{JAC-2000-MnN,JPSJ-1968-M2N,JCP-1955-Mn5N2} Mn$_3$N$_2$ forms the $\eta$ phase with a face-centered tetragonal structure, \cite{APL-2001-Yang} whereas Mn$_4$N has an antiperovskite crystal structure, which belongs to $\varepsilon$ phase. \cite{AIP-2020-Mn4N} The magnetism of these compounds can be determined by magnetic susceptibility or neutron diffraction measurements. MnN, Mn$_6$N$_5$, Mn$_2$N, and Mn$_3$N$_2$ all exhibit antiferromagnetism, \cite{JAC-2000-MnN,JPSJ-1968-M2N,JAC-1994-MnN} whereas Mn$_5$N$_2$ shows ferromagnetism, \cite{JCP-1955-Mn5N2} but Mn$_4$N is a ferrimagnet (see Figure 6b). \cite{IOP-2022-Mn4N,AIP-2020-Mn4N} Among these compounds, Mn$_4$N stands apart in terms of potential spintronic applications due to its room-temperature ferrimagnetism, strong perpendicular magnetic anisotropy, and low saturation magnetization. \cite{IOP-2022-Mn4N,AIP-2020-Mn4N}

The Fe-N system forms various phases depending on the nitrogen content, and the magnetic moments vary significantly among these phases. \cite{PRB-2010-FeN} The nitrogen-poor Fe-N phases are FM, including $\gamma'$-Fe$_4$N, $\varepsilon$-Fe$_{2-3}$N, $\alpha'$-Fe$_8$N and $\alpha ''$-Fe$_{16}$N$_2$. \cite{SurSci-2001-Grachev,PRB-2010-FeN,IEEE-1987-FeN,AM-2023-Fe3N} Figure 6c shows the temperature-dependent magnetization and crystal structure of FM Fe$_{3}$N. These nitrogen-poor phases attracted widespread research interest due to their high magnetic moments and tunable magnetic properties with varying concentrations of nitrogen. Specifically, the $\alpha ''$-Fe$_{16}$N$_2$ phase has been regarded as a promising material for high-density magnetic recording media since its magnetic moment is even higher than pure $\alpha$-Fe. \cite{JMMM-2000-FeN,JAC-2001-FeN,APL-2011-FeN,ASS-2003-FeN} With increasing nitrogen content, the ferromagnetism in Fe-N becomes weaker, as indicated by the very weak itinerant ferromagnetism in $\zeta$-Fe$_2$N. \cite{HI-1998-FeN} As for the FeN compound, the $\gamma'''$-NaCl-type phase is AFM while the $\gamma''$-ZnS-type phase is non-magnetic. \cite{HI-1998-FeN} Moreover, the Co-N system also possesses various phases. For example, Co$_2$N and CoN were reported to be Pauli paramagnets, \cite{EJIC-2016-CoN,JAC-1995-CoN} while Co$_{4+x}$N and Co$_3$N$_{1+y}$ were found to be ferromagnetic phases. \cite{JAC-2014-CoN}

The magnetism of the Cr-N system is not as complex as the above, the CrN compound was reported to be an antiferromagnet below its N$\acute{e}$el temperature $\sim$ 280 K, as shown in Figure 6d, accompanied by structure transition from cubic NaCl-type to orthorhombic structure. \cite{PRB-1999-CrN,PR-1960-CrN,PRL-2010-CrN,AM-2021-Jin} However, ferromagnetic-like behavior was observed in CrN thin films, \cite{APL-2006-Ney} and even room temperature ferromagnetism has been reported in CrN-Cr$_2$O$_3$ interfaces. \cite{PRL-2022-Jin}

Ternary nitride compounds in the form of Mn$_3$XN with antiperovskite structure are attracting rising interest, in which the Mn atoms form Kagome lattices and the coupled spins are frustrated. \cite{AM-2023-MnSnN,ACSAMI-2018-MnNiN,Acta-2014-MnCuN,NC-2020-MnGaN,PRB-2019-MnNiCuN,PRM-2019-MnNiN,AFM-2019-MnNiN,PRM-2024-MnGaN} An anomalous Hall effect, strain tunable magnetic order, and a giant piezomagnetism have been observed in the frustrated antiferromagnet Mn$_3$NiN. \cite{ACSAMI-2018-MnNiN,PRB-2019-MnNiCuN,PRM-2019-MnNiN,AFM-2019-MnNiN} The atomic displacements induced non-collinear antiferromagnetism have been observed in Mn$_3$SnN, leading to the realization of the anomalous Hall effect. \cite{AM-2023-MnSnN} The Hall resistance can be electrically manipulated by a small switching current in the noncollinear antiferromagnetic (AFM) Mn$_3$GaN/Pt bilayers at room temperature, highlights the important role of spin-torque effect. \cite{APL-2019-Hajiri} An unconventional spin-orbit torque (SOT) at room temperature has been demonstrated in Mn$_3$GaN/permalloy heterostructure,\cite{NC-2020-MnGaN} as shown in Figure 6e. Moreover, Hajiri et al. reported the SOT switching of Mn$_3$GaN in the absence of external magnetic field at room temperature.\cite{PRA-2021-Hajiri} All these findings facilitate the implementation of these magnetic TMN materials in antiferromagnetic spintronics. 

\subsection{Ferroelectricity}
Due to the excellent compatibility of nitride films with the semiconductor industry, ferroelectric nitrides have a promising future in microelectronics applications. Aluminum nitride (AlN), as an attractive piezoelectric material, has a significant and immediate impact in the 6G RF filter community and other related fields. Figure 7a-c illustrates the remarkable achievement of a 500\% increase in the piezoelectric coefficient with the introduction of transition element Sc doping in AlN (Sc$_{0.4} $ Al$_{0.6} $ N). \cite{AM-2009-AlScN} Notably, AlScN exhibits robust ferroelectric behavior, as reported in a study. \cite{JAP-2019-Ferr-AlScN}
\begin{figure}[b]
	\centering
	\includegraphics[width=0.7\linewidth]{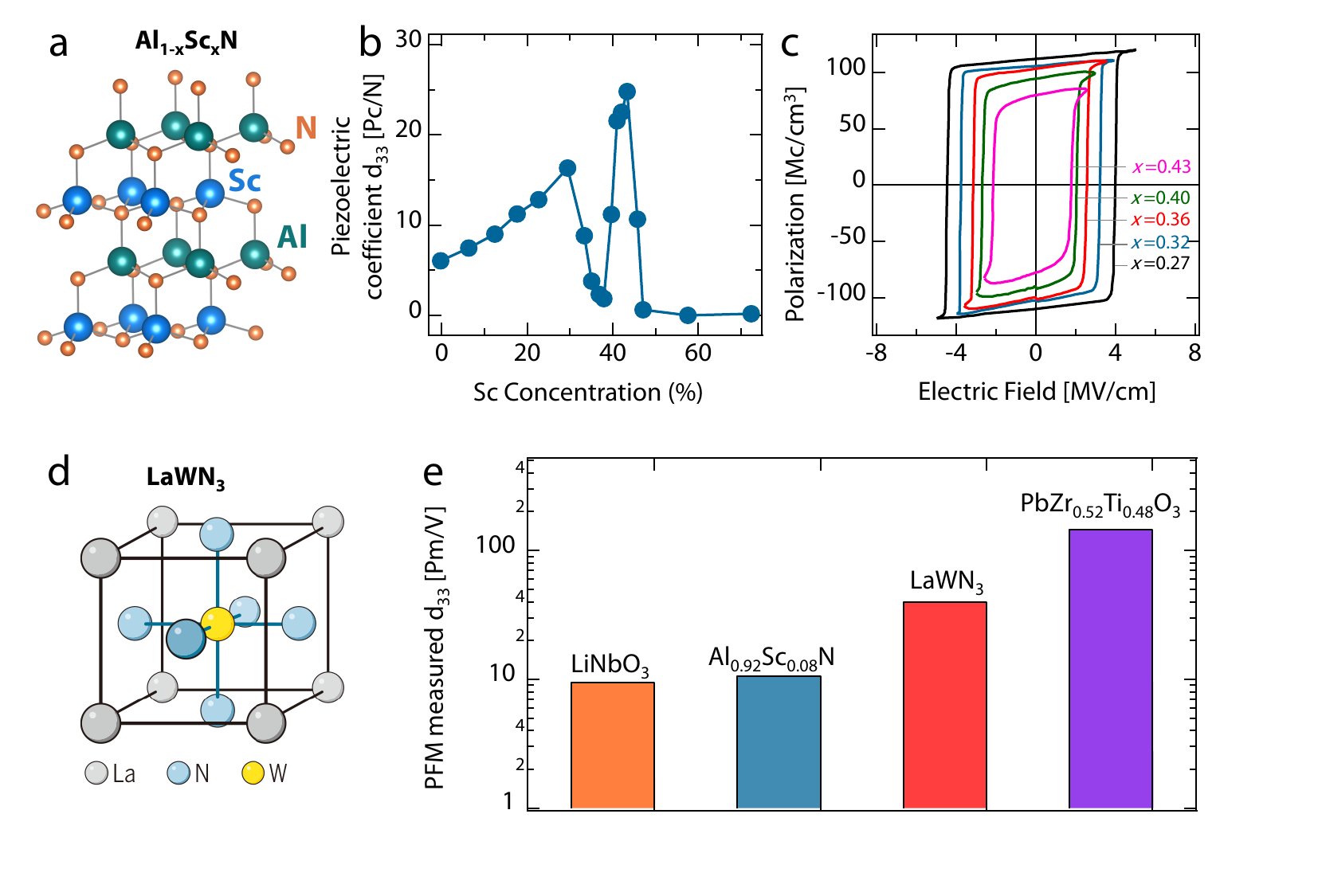}
	\caption{ a) Schematic of Al$_{1-x}$Sc$_x$N wurtzite structure. b) Piezoelectric coefficient $d_{33}$ \cite{AM-2009-AlScN} and c) P-E loops of ferroelectric Al$_{1-x}$Sc$_x$N with various Sc contents. \cite{JAP-2019-Ferr-AlScN} Reproduced with permission.\cite{AM-2009-AlScN}  Copyright 2009, Wiley-VCH. Reproduced with permission.\cite{JAP-2019-Ferr-AlScN}  Copyright 2019, AIP Publishing. d) Schematic of LaWN$ _3$ perovskite structure. \cite{Science-2021-LaWN-Polar} Reproduced with permission.\cite{Science-2021-LaWN-Polar} Copyright 2021, American Association for the Advancement of Science. e) Comparison of piezoelectric materials, demonstrating strong piezoelectric response of LaWN$ _3 $. \cite{Science-2021-LaWN} Reproduced with permission.\cite{Science-2021-LaWN} Copyright 2021, American Association for the Advancement of Science. }
\end{figure}

The search for stable nitride perovskites through high-throughput first principles calculations has resulted in the synthesis of lanthanum tungsten nitride (LaWN$ _3 $) as the only known ferroelectric perovskite nitride to date.  \cite{ChemMater-2015-Nitride,Science-2021-LaWN,Science-2021-LaWN-Polar} This achievement represents a pioneering demonstration of a polar nitride perovskite, enabling the integration of functional perovskites onto a semiconductor platform without the presence of oxygen. LaWN$ _3 $ exhibits a typical perovskite structure, as shown in Figure 7d. Notably, the piezoelectric coefficient d$ _{33} $ value of LaWN$ _3 $ is significantly higher than that of the piezoelectric oxide LiNbO$_3$ and the piezoelectric nitride Al$_{0.92}$Sc$_{0.08}$N, and is only surpassed by the value for PbZr$_{0.52}$Ti$_{0.48}$O$_{3}$, as shown in Figure 7e. The demonstration of oxygen-free nitride perovskite with a competitive piezoelectric response paves the way for integrating the rich functionalities of polar perovskites with the mainstream semiconductor industry. \cite{Science-2021-LaWN-Polar} 
However, the existence of ferroelectricity in LaWN$_3$ films remains inconclusive due to large leakage currents. \cite{Science-2021-LaWN,PRM-2023-LaWN} Therefore, it is necessary to obtain high-quality single crystalline LaWN$_3$ films not only to determine the presence of ferroelectricity but also to improve their properties.

\subsection{Plasmonics}

Plasmon, as collective oscillations of free electrons (see Figure 8a), has attracted tremendous interest due to its great capability of subwavelength confinement of electromagnetic waves and significantly enhanced electron-light interactions, which is widely applied in novel electronic/photonic devices, sensors, photochemistry applications, etc. \cite{AM-2013-BeyondGold,Science-2014-Refractory,AM-2014-Li,Science-2015-Gold,CM-2022-Dasog,ThinSolidFilms-1994-Steinmuller} TMNs (such as TiN, NbN, HfN, and ZrN) have garnered significant research interest as excellent plasmonic materials comparable to noble metals (e.g., Au and Ag). \cite{AM-2013-BeyondGold,Science-2014-Refractory,AM-2014-Li,Science-2015-Gold,ThinSolidFilms-1994-Steinmuller,CM-2022-Dasog} These TMNs exhibit superior stability in harsh environments, such as high temperatures and intensive light illumination, as well as excellent compatibility with complementary metal oxide semiconductor (CMOS) technology and biological systems. Moreover, their electronic and photonic properties can be finely tuned by stoichiometry and oxidation, outperforming noble metals and enabling a wide range of novel applications in plasmonics and nanophotonics. These applications include high-temperature sensors, flat optics, thermophotovoltaics, heat-assisted data recording, and thermal therapy, among others. \cite{ACSPho-2015-ENZ,ACSPho-2016-HotElectron,AOM-2017-TiNNano,ACSPho-2017-TiNLattice,ACSPho-2017-TiNNano,ACSInter-2017-TernaryNitrides,AOM-2017-Hot-ElectronTiN,ACSInter-2021-OLED,ACSInter-2021-Zhang} With advancements in epitaxy techniques, such as reactive sputtering, \cite{AOM-2017-UltrathinTiN} PLD, \cite{MaterToday-2021-Huang}, and nitrogen-plasma-assisted MBE, \cite{ACSPho-2019-Guo} high-quality epitaxial TMN plasmonic films with atomic-flat surface morphology and controllable thickness down to approximately 1 nm can be fabricated. This level of control is unattainable for typical noble metal films due to their high surface energy. These advancements have opened up new avenues in the study of hyperbolic metamaterials (HMMs), \cite{MaterToday-2021-Huang,PNAS-2014-TiN,ACSAEM-2022-HfZrN-ScN} and quantum-confinement-induced optics. \cite{AFM-2020-Quantum,SciAdv-2020-Quantum} As an archetype of TMNs, TiN has received more research attention compared to other TMNs, as shown in Figure 8c. 
 
\begin{figure}[h]
	\centering
	\includegraphics[width=0.95\linewidth]{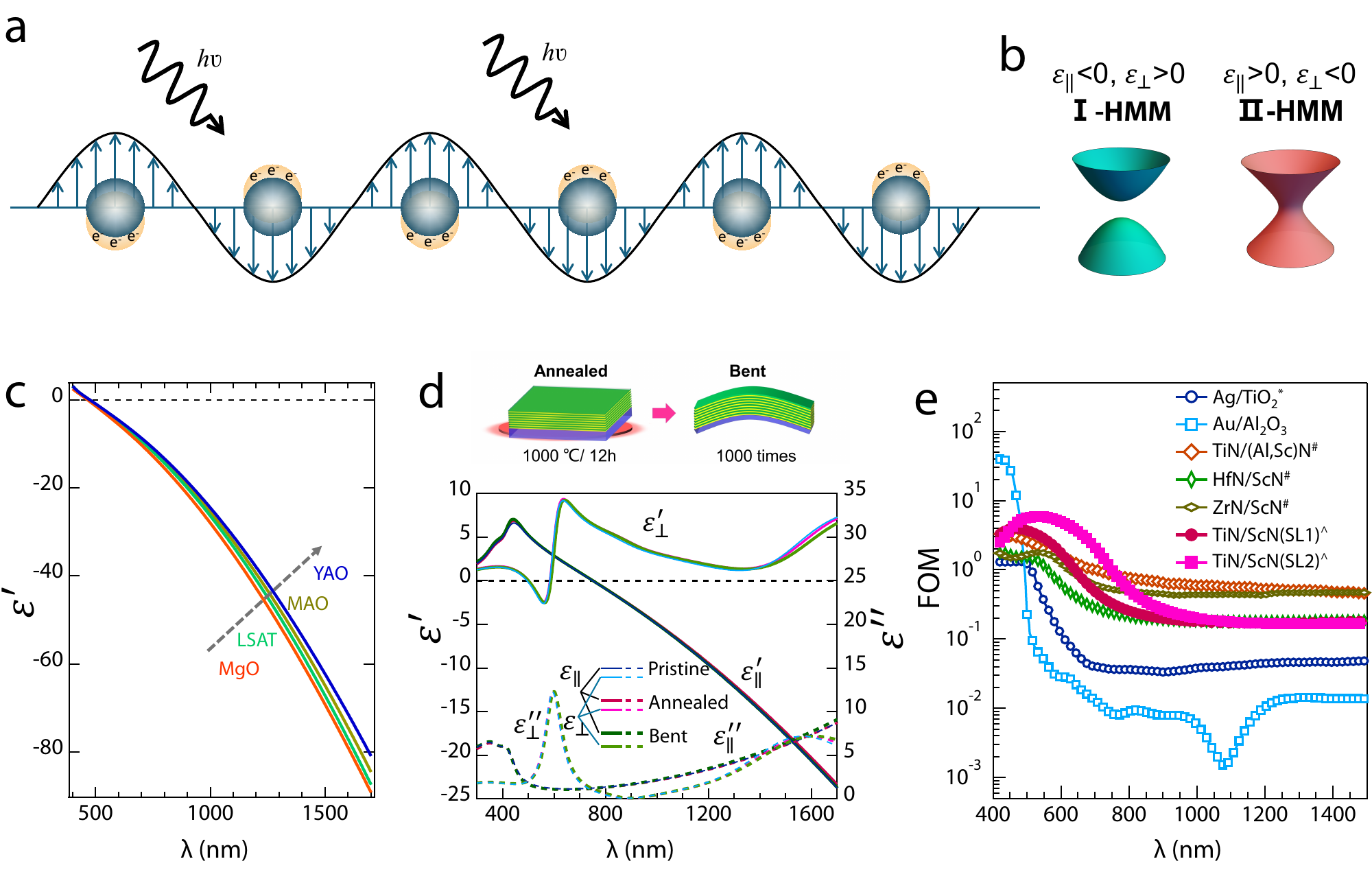}
	\caption{  a) Plasmonic properties. b) Isofrequency surfaces of extraordinary waves in type-\uppercase\expandafter{\romannumeral1} ($\varepsilon_{zz}$=$\varepsilon_{\parallel } $\textless0,$\varepsilon_{xx}$=$\varepsilon_{yy}$=$\varepsilon_{\bot}$\textgreater0) and type-\uppercase\expandafter{\romannumeral2} ($\varepsilon_{zz}$=$\varepsilon_{\parallel } $\textgreater0,$\varepsilon_{xx}$=$\varepsilon_{yy}$=$\varepsilon_{\bot}$\textless0) hyperbolic metamaterials. c) The real part $\varepsilon^{'}$ of TiN films on various substrates.\cite{PRM-2021-Bi} Reproduced with permission.\cite{PRM-2021-Bi} Copyright 2021, American Physical Society. d) Superior thermal and mechanical stability of flexible HMMs based on TiN/ScN superlattices. \cite{NanoLett-2023-Zhang} e) A comparison of figure of merit (FOMs) in existing HMMs based on rigid noble metal/dielectric multilayers (marked with *) and epitaxial TMN superlattices including rigid (marked with $\sharp$) and flexible ones (marked with $^\wedge$). \cite{NanoLett-2023-Zhang} d) and e) Reproduced with permission.\cite{NanoLett-2023-Zhang} Copyright 2023, American Chemical Society.}
\end{figure}

As optical metamaterials with extreme optical anisotropy are rarely found in nature, HMMs display hyperbolic dispersion (dielectric constants show opposite sign along one axis compared to the other two axes), i.e., type-\uppercase\expandafter{\romannumeral1} ($\varepsilon_{zz}$=$\varepsilon_{\parallel } $\textless0,$\varepsilon_{xx}$=$\varepsilon_{yy}$=$\varepsilon_{\bot}$\textgreater0) and type-\uppercase\expandafter{\romannumeral2} ($\varepsilon_{zz}$=$\varepsilon_{\parallel } $\textgreater0,$\varepsilon_{xx}$=$\varepsilon_{yy}$=$\varepsilon_{\bot}$\textless0) hyperbolic dispersion in Figure 8b). \cite{NP-2013-Hyperbolic,AOM-2019-Hyperbolic} HMMs have attracted growing interest in the last two decades due to their unique properties, including dramatic changes in the light propagation behaviors, peculiar emission properties, and diverging photonic density of states (PDOS), which enables many novel applications, including beam manipulation, super-resolution imaging, spontaneous and thermal emission engineering, ultrasensitive optical sensing, and broadband optical absorption. \cite{NP-2013-Hyperbolic,AOM-2019-Hyperbolic} 

In recent years, the successful epitaxy of high-quality TMN films made the fabrication of high-performance HMMs consisting of sing-crystalline, ultrathin TMN layers with atomic-smooth interface/surface possible, which outperform HMMs made of noble metal and dielectric multilayers. Conventional noble metals suffer from high surface energy, resulting in high roughness on their films. However, much lower surface energy enables much better control of the roughness at the interface/surface of TMN films, leading to much lower optical loss resulting from roughness. Fully epitaxial TMN-based HMMs were demonstrated by growing TiN/(Al,Sc)N superlattices with ultrathin TiN and (Al,Sc)N layers on lattice-matched MgO substrates, showing atomically sharp interface, excellent crystalline quality, and PODS enhancement over noble metal/dielectric based HMMs. \cite{PNAS-2014-TiN,PRM-2017-TiN-AlScN} Later, Huang et al. developed a method to integrate single-crystalline TiN with rock-salt oxide, i.e., MgO, to form hybrid HMMs by PLD. \cite{MaterToday-2021-Huang} It is remarkable that high-quality TiN/MgO superlattice with the thicknesses of TiN and MgO layers down to 0.9 nm and 0.63 nm, respectively, can be successfully fabricated, demonstrating type-\uppercase\expandafter{\romannumeral1} hyperbolic dispersion in visible ranges and type-\uppercase\expandafter{\romannumeral2} hyperbolic dispersion in NIR ranges. With the rise of flexible plasmonics and photonics, Zhang et al. successfully demonstrated a scalable route to fabricating flexible, high-performance, and refractory HMMs by directly depositing epitaxial TiN/ScN superlattices on flexible inorganic fluorophlogopite-mica (see Figure 3b), a synthetic two-dimensional van der Waals layered material with thermal stability up to 1100 $^\circ$C, via reactive sputtering. \cite{NanoLett-2023-Zhang} Excitingly, these flexible single-crystalline HMMs based on TiN/ScN superlattices show dual-band hyperbolic dispersion and remarkable stability during 1000 $^\circ$C heating or after being bent 1000 times (see Figure 8d). Furthermore, by comparing the figure of merits (FOMs) defined as Re($k_\bot $)/Im($k_\bot $) (Re($k_\bot $) and Im($k_\bot $) are the real and imaginary part of the propagation components perpendicular to the film plane) in existing HMMs, it can be concluded that HMMs based on epitaxial TMN superlattices, regardless of rigid (marked with $\sharp$) and flexible ones (marked with $^\wedge$), have superior performance than those in HMMs based on rigid noble metal/dielectric multilayers (marked with *), as shown in Figure 8e.

\section{Advanced Devices}

Superconducting TMNs have emerged as promising materials for key devices in superconducting quantum computers, including the superconducting qubits\cite{CM-2021-Qubit} and  superconducting Coplanar-Waveguide Resonators, \cite{PRA-2019-SCWR} as shown in Figure 9a,b. Dielectric loss is one of the major decoherence sources of superconducting qubits, mainly originating from the surfaces and interfaces of the superconducting films and the substrates.\cite{PRA-2023-Deng} Meanwhile, the coherence times of superconducting quantum circuits made by conventional Al-based JJs are limited by energy or phase relaxation due to microscopic two-level systems (TLSs) in the amorphous aluminum oxide (AlO$_x$) tunnel barriers.\cite{CM-2021-Qubit,PRL-2005-Martinis,RPP-2019-Muller,IEEE-2009-McDermott} Considerable effort has been invested in the search for superior superconducting materials to enhance superconducting quantum circuits.\cite{Science-2021-Material,MRS-2013-Oliver} TiN has been proven to have ultrahigh quality factors, long qubit coherence time, high kinetic inductance, and excellent chemical stability for advanced  superconducting quantum circuit fabrication.\cite{JAP-2012-Krockenberger, APL-2010-Vissers,APL-2013-Chang,APL-2014-Shearrow,APL-2020-Melville, APL-2024-Zhang} The Alibaba Quantum Laboratory in China concentrated on the high-quality superconducting TiN thin films in the superconducting quantum devices. The highly crystalline TiN films superconducting qubits show lifetimes of up to 300 $\mu$s and quality factors achieving 8.1 million,\cite{PRA-2023-Deng} and the single-photon quality factor of the epitaxial TiN microwave resonators is as high as 3.3 $\times$ 10$^6$.\cite{PRM-2022-Gao} Furthermore, the all-nitride qubits (NbN/AlN/NbN epitaxial
Josephson junctions) were synthesized by the National Institute of Information and Communications Technology (NICT) of Japan, which demonstrated a significant improvement in coherence times ($\bar{T_1}=16.3 \mu s $ and $\bar{T_2}=21.5 \mu s $).\cite{CM-2021-Qubit} The all-nitride qubits have great advantages such as chemical stability against oxidation, resulting in fewer two-level fluctuators, feasibility for epitaxial tunnel barriers that reduce energy relaxation and dephasing, and a larger superconducting gap of $\sim$ 5.2 meV for NbN, compared to $\sim$ 0.3 meV for Al, which suppresses the excitation of quasiparticles.\cite{CM-2021-Qubit}  Moreover, the superconducting TMNs with high kinetic inductance are suitable for the development of noise-resilient superconducting qubits and high-impedance circuits,\cite{AM-2022-Gao} and as Microwave Kinetic Inductance Detectors (MKIDs) are crucial for space exploration.\cite{JLTP-2018-Morozov,PRA-2023-Kouwenhoven,APL-2023-Boussaha,OptExpress-2012-Mazin} On the other hand, superconducting detectors, with their interaction between superconducting macroscopic quantum effects and the environmental electromagnetic field, can provide ultra-sensitive detection with quantum-limit sensitivity. Superconducting nanowire single-photon detectors (SNSPDs, as shown in Figure 9c)  play a crucial  role in quantum information, deep space communication, and lidar.\cite{PHYSICS-2021-SNSPD} It is noted that NbN and NbTiN are among the most common materials in SNSPDs. \cite{APL-2021-Zadeh}

Furthermore, significant progress would be expected by combining the advantages of quantum technology with the power of semiconductors (see Figure 9d). \cite{SciAdv-2021-NbN-AlN, Nature-2018-AlN-NbN,IEEE-2016-NbN-GaN,IEEE-2018-NbN-GaN} Additionally, the typically incompatible integer quantum Hall effect and superconductivity was found to exist concurrently in epitaxial GaN two-dimensional electron gas / NbN heterostructures. \cite{SciAdv-2021-NbN-AlN}  The NbN$
_x$-based singlephoton detectors can be integrated with GaN HEMT amplifiers for secure quantum communications. Moreover, combining GaN HEMT microwave amplifiers with NbN$
_x$-based Josephson junctions can provide an all-epitaxial platform for superconducting qubits.\cite{Nature-2018-AlN-NbN} Epitaxial integration of semiconductor/ superconductor heterostructures enables the integration of macroscopic quantum effects from superconductors with the electronic, photonic, and piezoelectric properties of the group \uppercase\expandafter{\romannumeral3}/nitride semiconductor family, \cite{Nature-2018-AlN-NbN} which is highly desirable for future quantum computation. On the other hand, the excellent  electrical conductivity of TMN films is commonly used as diffusion barrier layers and gate electrodes in semiconductor technology (see Figure 9e). \cite{Science-2023-HZO,SciRep-2014-TiN}
\begin{figure}[t]
	\centering
	\includegraphics[width=0.95\linewidth]{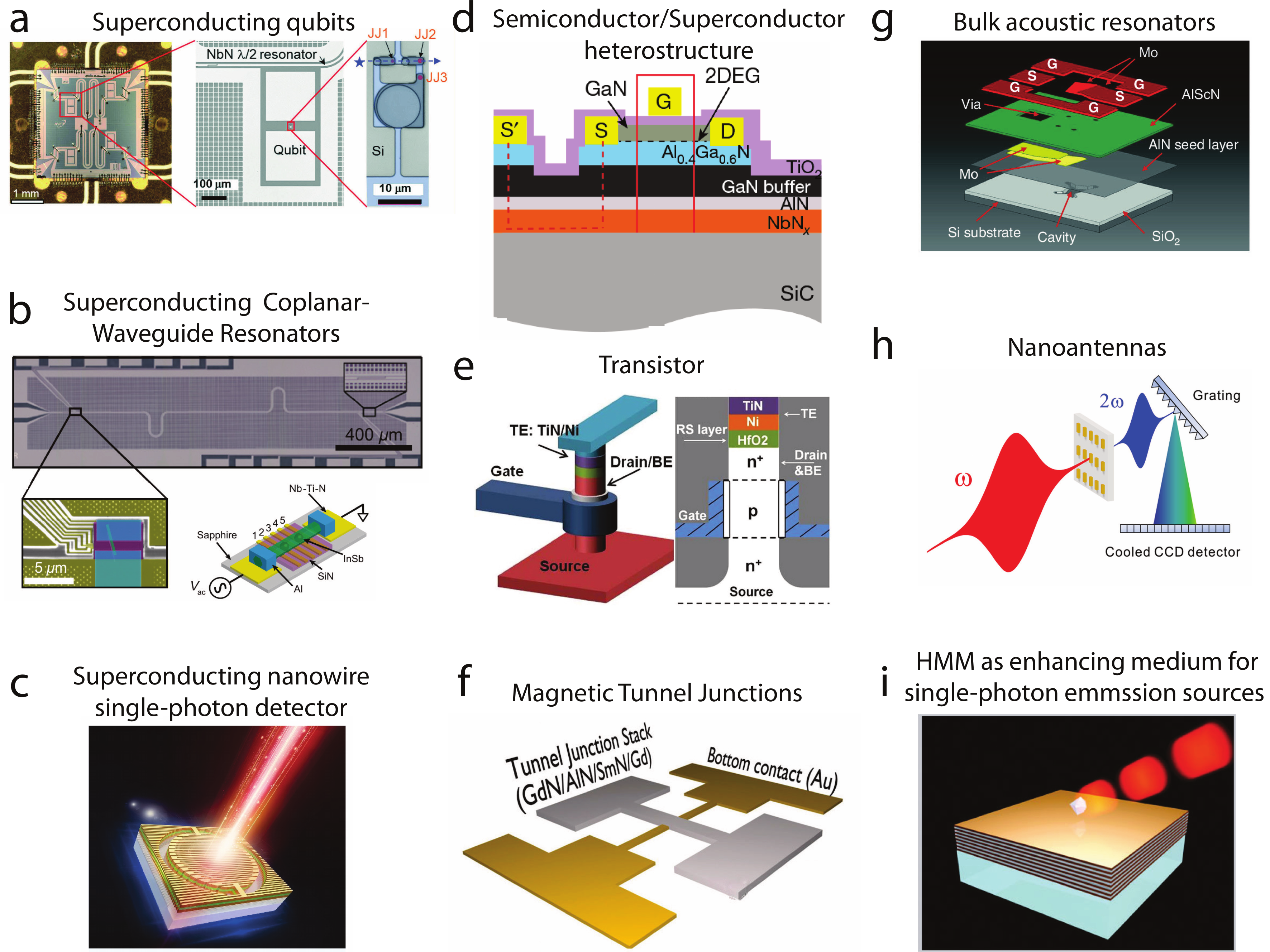}
	\caption{a)  Superconducting qubits. \cite{CM-2021-Qubit} Reproduced with permission.\cite{CM-2021-Qubit} Copyright 2021, Nature Publishing Group UK London. b) Superconducting  Coplanar-Waveguide Resonators. \cite{PRA-2019-SCWR} Reproduced with permission.\cite{PRA-2019-SCWR} Copyright 2019, American Physical Society. c) Superconducting nanowire single-photon detector. \cite{PHYSICS-2021-SNSPD} Reproduced with permission.\cite{PHYSICS-2021-SNSPD}  Copyright 2021, Chinese Physical Society. d) Semiconductor/superconductor heterostructure. \cite{Nature-2018-AlN-NbN} Reproduced with permission.\cite{Nature-2018-AlN-NbN}  Copyright 2018, Nature Publishing Group UK London. e) Transistor. \cite{SciRep-2014-TiN} Reproduced with permission. \cite{SciRep-2014-TiN} Copyright 2014, Nature Publishing Group UK London. f) Magnetic Tunnel Junctions. \cite{PRA-2016-Magnetic} Reproduced with permission. \cite{PRA-2016-Magnetic} Copyright 2016, American Physical Society.  g) Bulk acoustic resonators. \cite{MN-2022-AlN} Reproduced with permission.\cite{MN-2022-AlN}  Copyright 2022, Nature Publishing Group UK London. h) Nanoantennas. \cite{NanoLett-2016-Nonlinear} Reproduced with permission.\cite{NanoLett-2016-Nonlinear} Copyright 2016, American Chemical Society. i) HMM as enhancing medium for single-photon emmssion. \cite{Laser-2015-Metamaterial} Reproduced with permission.\cite{Laser-2015-Metamaterial}  Copyright 2015,Wiley-VCH }
\end{figure}

Magnetic TMNs have great potential prospects in magnetic tunnel junction (MTJs) devices  due to their thermal stability, low coercivity, and low magnetization.
\cite{JAP-2010-Kim,JJAP-2007-Narahara} The magneto-resistive random-access memory (MRAM), which utilizes the tunnel magnetoresistance (TMR) effect to achieve the nonvolatile memory function. The TMR effect is MTJs devices, which composed of two ferromagnetic electrodes separated by a thin insulating layer, as illustrated in Figure 9f. \cite{PRA-2016-Magnetic}  Kim et al. developed an all-nitride MTJ of CoFeN/AlN/CoFeN by using CoFeN as electrodes and AlN$_x$ as tunneling barriers.\cite{JAP-2010-Kim} They found that the introduction of a small amount of nitrogen into the CoFe film is the key to control the nanocrystalline phases in the CoFeN film and leads to films with low coercivity and magnetization values, which are crucial for a low switching field in MTJs. A similar all-nitride MTJ of Fe$_3$N/AlN/Fe$_4$N has been fabricated by conventional molecular beam epitaxy.\cite{JJAP-2007-Narahara} Clear two-step hysteresis loops were observed in the magnetization versus magnetic field curves, corresponding to the two distinct nitride ferromagnetic layers.  Some of the magnetic TMNs mentioned above, such as Fe-N and Mn-N compounds, show potential applications in spintronic devices due to their peculiar magnetic properties. For example, a current-perpendicular-to-plane (CPP) giant magnetoresistance (GMR) device has been fabricated based on the low-damping, and non-interface perpendicular Fe$_{16}$N$_2$ thin film.\cite{PPL-2019-Li} The current-induced magnetization switching in Au/Fe$_4$N bilayer films\cite{ACSAMI-2019-Li} and the large current-driven domain wall mobility in Mn$_4$N thin films\cite{NanoLett-2019-Mn4N} demonstrate the manipulation of magnetic properties by charge current, which is essential for future spintronic applications.  The spin transfer torque (STT) is another type of high-speed and nonvolatile MRAM storage where the magnetic layer plays a key role in their capacity. The magnetic properties of Mn$_4$N films, including ferrimagnetism, perpendicular magnetic anisotropy (PMA), low saturation magnetization, and the tunable critical switching behavior upon doping, are promising for the development of STT devices.\cite{JAP-2014-Yasutomi, JPDAP-2021-Zhang} The modification of anomalous Hall effect (AHE) in  Mn$_4$N/Au bilayer induced by interfacial spin-orbital coupling from the heavy metal provides potential strategy to develop spintronic devices.\cite{JAP-2018-Wang} Moreover, the spin states in Mn$_4$N films are ferroelectrically tunable as revealed by the observation of a topological Hall effect in Mn$_4$N/PMN-PT heterostructure, paving a path to non-magnetic manipulation method and related device development.\cite{APL-2018-Wang} The emerging magneto-ionics, which is based on the voltage-driven ion transport in magnetic materials, opens another opportunity to utilize magnetic TMNs in advanced spintronic devices. Magneto-ionics is a branch of voltage-controlled magnetism that injects ions such as O$^{2-}$, H$^{+}$, Li$^{+}$, F$^{-}$, or N$^{2-/3-}$ into target material by applying voltage instead of electric current to achieve modulations of magnetic properties. Compared to other ions, nitrogen magneto-ionics requires lower threshold voltages and exhibits improved rates and cyclability, owing to the lower activation energy for ion diffusion and the lower electronegativity of nitrogen.\cite{NC-2020-Rojas} The voltage-driven nitrogen ion motion has been demonstrated in CoN, FeN, and CoFeN films.\cite{NC-2020-Rojas,ACSAMI-2021-Rojas, AEM-2023-Monalisha} In particular, a CoFeN based synapse device has achieved distinct multilevel non-volatile magnetic states for analog computing and multi-state storage.\cite{AEM-2023-Monalisha} As demonstrated by its remarkable synapse retention time, high switching ratio, and large endurance, nitride-based magneto-ionic devices provide a new approach to implement neuromorphic computing.\cite{AEM-2023-Monalisha}

Ferroelectric  Al$_{1-x}$Sc$_x$N is an emerging material in  telecommunications technology. The emergence of 5G, Wi-Fi, and 4G LTE communications is driving the demand for filters with higher frequencies and wider bandwidths. Microelectromechanical system (MEMS) filters, such as surface acoustic wave (SAW) and bulk acoustic wave (BAW) filters, hold promise as candidates for operation in new radio frequency (RF) bands. It is challenging for a traditional aluminum nitride (AlN)-based BAW filter to meet several allocated 5G bands.\cite{MN-2022-AlN} Al$_{1-x}$Sc$_x$N  exhibits robust ferroelectric behavior and shows an increase in the piezoelectric coefficient compared to the AlN,\cite{JAP-2019-Ferr-AlScN,AM-2009-AlScN}  the Al$_{1-x}$Sc$_x$N-based film bulk acoustic wave resonator (FBAR, see Figure 9g) opens avenues for wideband acoustic filters to operate within the 5G band. \cite{MN-2022-AlN,Vac-2011-AlScN,JMS-2020-AlScN}

In plasmonics, TMNs are ideal nonlinear materials compared to noble metals due to their high-temperature sustainability and large optical nonlinearities. As seen in Figure 9h, TiN nanoantenna arrays demonstrate superior second-harmonic generation efficiency and durability upon strong laser illumination (average power up to 200 mW at 44 MHz repetition rate, and peak intensity up to 15 GW/cm$_2$), an order of magnitude higher incident energy than gold for a similar laser pulse.\cite{NanoLett-2016-Nonlinear} In sharp contrast to TiN nanoantenna arrays, the gold nanoantenna arrays show severe nanostructure degradation upon intensive light illumination.\cite{NanoLett-2016-Nonlinear} The TMN superlattices (e.g., TiN/(Al,Sc)N superlattices) based HMMs can be used to enhance the single–photon emission from nitrogen–vacancy color centers in nanodiamonds by providing more PDOS and decay channels for the coupled single-photon emitters (Figure 9i), \cite{Laser-2015-Metamaterial} which is very promising for building highly efficient room temperature CMOS-compatible single-photon sources. In addition to nanoantennas and enhancing medium for single-photon emission sources, single-crystalline TMN films and heterostructures can be applied in many other nonlinearity devices, thermal plasmonic devices, and quantum optical devices.

\section{Summary and Outlook}

Compared to previous reviews in the fields of catalysis, coatings, and plasmonics, this review offers a brief overview of recent advancements in epitaxial transition-metal nitride films from the perspective of materials physics and condensed matter physics. The focus is on summarizing their physical properties, e.g., superconductivity, magnetism, and ferroelectricity, as well as their applications in advanced electronic and plasmonic devices. Understanding the physics behind these materials requires advanced deposition techniques, such as MBE, PLD, and magnetron sputtering epitaxy, capable of synthesizing high-quality epitaxial transition-metal nitride films at the atomic scale. In this review, we highlight the development of transition-metal nitirde films can be as well as transition-metal oxide films and transition-metal dichalcogenides in near future. To advance the field of high-performance electronics and plasmonics, it is crucial to develop wafer-scale transition-metal nitride films, up to 6 inches in size, with exceptional purity levels, reaching up to 99.9999\%. Among the various deposition methods, magnetron sputtering epitaxy shows promise as an effective approach to meet this objective. We suggest that the synthesis of high-quality, wafer-scale, and uniform epitaxial transition metal nitride films with atomic precision is achievable through the introduction of the high-pressure magnetron sputtering epitaxy. Only with high-quality single-crystalline samples it is possible to reveal the hidden electronic structures of TMNs using advanced techniques such as Scanning Tunneling Microscopy (STM) and Angle-resolved Photoemission Electron Spectroscopy (ARPES).

\section{Acknowledgments}

We thank Professor Tao Chen for fruitful discussions and acknowledge Wei Chen and Rongjing Zhai for their valuable assistance in plotting figures. This work was supported by the National Key Research and Development Program of China (Grant Nos. 2024YFF0508500 and 2022YFA1403000), the Zhejiang Provincial Natural Science Foundation of China, the “Pioneer” and “Leading Goose” R\&D Program of Zhejiang (2024C01252(SD2)), and the Ningbo Science and Technology Bureau (Grant No. 2022Z086).

\newpage

\end{document}